\documentclass[aps,reprint,nofootinbib,prd]{revtex4-1}

\pdfoutput=1

\usepackage[utf8]{inputenc}
\usepackage[T1]{fontenc}
\usepackage{mathpazo}
\usepackage{graphicx}
\usepackage{hyperref}
\usepackage{amsmath}
\usepackage{colonequals}
\usepackage{xcolor}

\definecolor{darkred}{rgb}{0.9,0,0}
\definecolor{darkgreen}{rgb}{0,0.9,0}
\definecolor{darkblue}{rgb}{0,0,0.9}

\hypersetup{
colorlinks,
linkcolor=darkblue,
filecolor=darkgreen,
urlcolor=darkblue,
citecolor=blue}

\begin{document}

\title{Graphene-based detectors for directional dark matter detection}\thanks{This is a preprint of an article published in European Physical Journal C. The final authenticated version is available online at \url{https://doi.org/10.1140/epjc/s10052-019-7071-2}.}
\author{Shang-Yung Wang}
\affiliation{Department of Physics, Tamkang University, Tamsui District, New Taipei City 25137, Taiwan}

\begin{abstract}
Dark matter detectors with directional sensitivity have the capability to distinguish dark matter induced nuclear recoils from isotropic backgrounds, thus providing a smoking gun signature for dark matter in the Galactic halo. Motivated by recent progress in graphene and two-dimensional materials research, we propose a novel class of directional dark matter detectors utilizing graphene-based van der Waals heterostructures. A conceptual design of the detector based on graphene/hexagonal boron nitride and graphene/molybdenum disulfide heterostructures is developed and analyzed. The proposed detector has modular scalability, keV-scale detection threshold, nanometer position resolution, sensitivity down to 10~$\mathrm{GeV}/c^2$ dark matter mass, and intrinsic head-tail discrimination and background rejection capabilities.
\end{abstract}


\maketitle

\section{Introduction}

A wide range of independent astrophysical observations on galactic and cosmological scales strongly indicate that more than 80\% of the matter in our Universe is in a form of nonluminous, nonbaryonic dark matter (DM)~\cite{Rubin:1970zza,Rubin:1980zd,Tyson:1998vp,Refregier:2003ct,Allen:2002eu,Bennett:2012zja,Ade:2013zuv}. 
The nature of DM remains to date a deep mystery, and its identification is one of the major outstanding problems in both astrophysics and particle physics. Many theoretical extensions of the standard model of particle physics predict new types of stable, charge neutral particles that could be candidates for DM~\cite{Bertone:2004pz,Feng:2010gw,Bertone:2018krk}. 
At present the leading DM candidates are weakly interacting massive particles (WIMPs). They are predicted to have mass in the range ${\sim\!1}~\mathrm{GeV}/c^2$--1~$\mathrm{TeV}/c^2$ and couple only weakly to ordinary standard model particles. 
Theoretical studies show that galactic WIMPs have promising prospects for direct detection~\cite{Bergstrom:2012fi}.

If WIMPs constitute the DM halo of our Galaxy, which is assumed to be nonrotaing in the Galactic rest frame, the fact our Solar System revolves around the Galactic center implies that halo WIMPs in the vicinity of the Earth may be directly detectable through their elastic scattering off atomic nuclei in terrestrial detectors~\cite{Goodman:1984dc}. 
In particular, the observed nuclear recoil events are expected to exhibit annual modulation as a result of Earth's orbital motion around the Sun~\cite{Drukier:1986tm,Freese:1987wu,Freese:2012xd}. 
For decades numerous experiments have been designed for direct detection of WIMP DM~\cite{Baudis:2012ig,Undagoitia:2015gya}. 
The DAMA/LIBRA experiment has recently reported observations of annual modulation with high statistical significance~\cite{Bernabei:2013xsa,Bernabei:2018yyw}. The modulation is consistent with a WIMP signal, but seems not compatible with the null results from the competing LUX~\cite{Akerib:2016vxi} and XENON~\cite{Aprile:2017iyp,Aprile:2018dbl} experiments.
While there is so far no conclusive evidence for WIMP detection, experimental limits on the elastic WIMP-nucleon scattering cross section for a wide range of WIMP masses have been obtained~\cite{Akerib:2016vxi,Aprile:2017iyp,Aprile:2018dbl}. 
The situation at the moment is more intriguing than perplexing and calls for further experimental and theoretical investigations.

The motion of the Earth and Sun relative to the Galactic center also gives rise to a strong directional dependence in WIMP-induced nuclear recoils~\cite{Spergel:1987kx}. 
This is because nuclear recoils resulting from WIMP collisions will significantly peak in the direction of the WIMP flux, producing a large forward-backward asymmetry. 
The WIMP flux direction is opposite to the direction of the Sun's motion around the Galactic center, which at the current epoch points towards the constellation Cygnus. 
On the contrary, nuclear recoils caused by random backgrounds (e.g., alphas, electrons, neutrons, neutrinos, gammas, and cosmic rays) are expected to be isotropic and exhibit no preferred direction. 
Detailed studies~\cite{Copi:1999pw,Morgan:2004ys,Billard:2012qu} show that the strong directional dependence is a robust signal of Galactic halo WIMPs in that it is generic to a wide variety of halo models, WIMP parameters, detector parameters, etc. 
In particular, for detectors with fine directional resolution if the directions of the recoils (head-tail directionality) are known then only of order ten recoil events will be sufficient to distinguish a WIMP signal from isotropic backgrounds, with the uncertainties in reconstructing the recoil direction only mildly increasing the required number of events~\cite{Morgan:2004ys}. 
Therefore, directional signal in nuclear recoil events serves as a ``smoking gun'' signature for direct detection of halo WIMPs. 
Various directional DM detection techniques have been proposed, and there have been increasing experimental efforts devoted to directional DM detection (for a recent review see, e.g., Ref.~\cite{Mayet:2016zxu}). The most notable ongoing experiments are DMTPC~\cite{Ahlen:2010ub}, DRIFT~\cite{Battat:2014van}, NEWAGE~\cite{Nakamura:2015iza}, and MIMAC~\cite{Riffard:2015rga}. 
All of those experiments utilize low-pressure gaseous time projection chambers (TPCs) to reconstruct the tracks of recoiling nuclei~\cite{Ahlen:2009ev}. 

In this article, we analyze the widely used TPC-based and the recently proposed DNA-based directional detectors~\cite{Drukier:2012hj,Drukier:2014rea}. 
We find that the design of the DNA-based detectors has remarkable merits that may hold the key to a new generation of detection technology with nanometer precision.
Motivated by those merits and recent progress in nanotechnology and graphene-related research, we propose a novel class of directional DM detectors using graphene-based heterostructures that are within the grasp of current technology. 
The proposed detector has modular scalability, keV-scale detection threshold, nanometer position resolution, sensitivity down to DM of ${\sim\!10}~\mathrm{GeV}/c^2$ mass, and intrinsic head-tail discrimination and background rejection capabilities.
The feasibility of this conceptual design is strongly supported by a great amount of experimental evidence on properties and fabrication of the heterostructures, and by simulation results.

The rest of this article is organized as follows. In Sec.~\ref{sec:DNA2D}, we highlight the design merits of the DNA-based directional detectors, and argue that graphene-based heterostructures are far more better than DNA arrays as the detection material for a new class of directional detectors with nanometer precision. In Sec.~\ref{sec:design},
we present the conceptual design, detection principle, and directionality and background rejection capabilities of the proposed graphene-based directional detector. Finally, we conclude in Sec.~\ref{sec:conclusion}.

\section{From DNA- to Graphene-based Directional Detectors}\label{sec:DNA2D}

TPCs provide three-dimensional information about the tracks of charged particles that traverse the chamber and ionize the detection material along the way. While dual phase (liquid/gas) argon and xenon TPCs have long been used in the DarkSide, LUX, PandaX, and XENON experiments~\cite{Undagoitia:2015gya}, they were not designed to provide recoil track information. 

In the current low-pressure gaseous TPC-based directional detection experiments~\cite{Ahlen:2010ub,Battat:2014van,Nakamura:2015iza,Riffard:2015rga}, the active volume of low-pressure gaseous TPCs is limited to ${\sim\!1}~\mathrm{m}^3$ or less and the target gas ($\mathrm{CS_2}$, $\mathrm{CF_4}$, $\mathrm{^3 He}$, or mixture thereof) is pumped to a low-pressure at ${\sim\!40}$--150~Torr~\cite{Ahlen:2009ev}. 
The purpose of a low-pressure, small-volume design is to ensure the tracks of the low-energy recoiling nuclei (with a typical recoil energy not more than a few tens of keV~\cite{Undagoitia:2015gya}) to be long enough for adequate reconstruction.
Since the gas also acts as the target material, a dilute target gas would inevitably imply a low event rate and a limited scalability.
Moreover, the information about the recoil tracks encoded in the drifting ionization electrons and/or ions is read out by a multiwire proportional counter, charge-coupled device, micropixel chamber, micromesh gaseous structure, or gas electron multipliers~\cite{Ahlen:2009ev}. 
The state-of-the-art position resolution of these readout techniques is of order ${\sim\!10}~\mu$m~\cite{Mounir:2015}, which however falls short of the fine directional resolution required for determining directional dependence. 
To achieve an unambiguous detection of DM, the existing TPC-based directional detectors would need to be improved for better scalability, sensitivity, and position resolution. 
Several promising ideas that independently exploit columnar recombination in a high-pressure xenon gas TPC~\cite{Nygren:2013nda,Gehman:2013mra}, capitalize on a combined optical readout of low noise complementary metal-oxide-semiconductor sensors and fast photomultiplier tubes~\cite{Antochi:2018otx}, and utilize a negative ion TPC operated at nearly atmospheric pressure (610~Torr)~\cite{Baracchini:2017ysg} have recently been suggested to address the issues of scaling-up feasibility and detection sensitivity. 
Nevertheless, with so much at stake, exploration of the possibility of new directional detection technology is clearly warranted.

Many new directional detection techniques using biological materials~\cite{Drukier:2012hj,Drukier:2014rea}, anisotropic scintillators~\cite{Bernabei:2003ct,Belli:1992zb,Cappella:2013rua,Rajendran:2017ynw}, ancient minerals~\cite{Drukier:2018pdy}, two-dimensional materials~\cite{Hochberg:2016ntt,Capparelli:2014lua,Cavoto:2016lqo,Cavoto:2017otc}, and electromagnetic filters~\cite{Baracchini:2018wwj,Betti:2019ouf,Betti:2018bjv} have been proposed. 
Here we focus on the interesting idea of biological DM detectors~\cite{Drukier:2012hj,Drukier:2014rea} that is closely relevant to the proposed graphene-based detector. Of particular novelty is the use of DNA in lieu of more conventional detection materials to provide high-resolution directional detection of DM. 
The basic detector unit consists of a thin film of metal (e.g., gold, tungsten, tin, or copper), from which a large number of single-stranded DNA (ssDNA) strands hang down.
The detector is modular in that a series of independent basic units (or modules) may be stacked on top of each other. 
The sequence of the ssDNA strands is known. When a WIMP scatters elastically off a nucleus in, say, a gold film, the recoiling nucleus traverses and, whenever it hits one, breaks a few hundreds of ssDNA strands before either coming to a stop or being captured in the Mylar (polyethylene terephthalate) film on the other side of the ssDNA layer. 
The fragmented strands can be recovered and removed periodically and then sequenced with a single base precision, corresponding to a precision of 0.7~nm for straightened ssDNA strands. Thus the path of the recoiling nucleus can be tracked with nanometer precision. 

Since the concept of DNA-based detectors is fundamentally distinct from that of the conventional TPC-based ones, it is therefore insightful to compare the two design concepts.
In fact, a thorough comparison reveals that DNA-based directional detectors have several remarkable merits that may hold the key to new directional detection technology. 

The first merit is dual material design, i.e., two different materials, instead of a single one, are separately used for target and detection. In this dual material design there is a constructive division of labor.
A dense high-atomic-number (high-$Z$) material (e.g., metal or liquid noble gas) is ideal for achieving a high event rate, while a sparse low-$Z$ material (e.g., soft or organic matter) is well suited for tracking recoiling nuclei.
The second merit is modularity in design, i.e., a large-size detector can be built up by stacking on top of each other, and/or placing side by side, a large number of independent modules. Evidently, modularity in design implies scalability. 
Importantly, we note that the independent modules may use a wide variety of target and detection material pairs which are sensitive to WIMPs in different mass ranges. 
This feature not only increases the nuclear recoil event rate but also allows WIMPs of different masses to be detected by a single detector.
The third merit is direct information encoding and delayed information decoding. Information about the recoil tracks is encoded directly in the detection material and decoded at a later time by a readout device that is \emph{not} an integral part of the detector. 
This information processing is in sharp contrast to that of the TPC-based detectors, in which track information is encoded in the fleeting excitations and/or relaxations (e.g., photons, phonons, ionization electrons and/or ions) produced in the detection material and hence has to be decoded in real time by an integral readout device.
The advantages in cost and scaling-up are quite evident since only a single readout device is needed regardless of the size of the detector.
The fourth, and the most important, merit is the \emph{intrinsic} nanometer structure of the detection material. Recall that for TPC-based detectors, the tracking precision is prone to diffusion of ionization electrons and/or ions during drift and is limited to ${\sim\!10}~\mu$m by the position resolution of readout devices. 
In light of the direct information encoding and delayed information decoding, nuclear recoil tracks can be measured with nanometer precision by a suitable combination of detection materials with an intrinsic nanometer structure and readout devices with a commensurate spatial resolution.

With these merits in mind, we now argue that graphene-based heterostructures are ideally suited as the detection material of a new class of directional detectors with nanometer precision.
Graphene is a two-dimensional (2D) atomic crystal which consists of carbon atoms tightly packed into a hexagonal lattice. Since its first isolation by mechanical exfoliation in 2004~\cite{Novoselov:2004}, graphene has been attracting tremendous interest and under intense experimental and theoretical study. 
Graphene has many extraordinary electronic and mechanical properties, such as high electron mobility, high thermal conductivity, excellent elastic flexibility, complete impermeability to any gases, and stability under ambient conditions~\cite{Novoselov:2012}. 
Recently, growth of bilayer-free, high crystalline quality monolayer graphene by chemical vapor deposition (CVD)~\cite{Li:2009} on low-cost copper films and its clean transfer, free of metal contaminants, structural defects, and polymeric residues have been achieved~\cite{Deokara:2015}. 
More recently, large-area high-quality graphene films of dimensions $3\times 3~\mathrm{cm}$ have been grown by CVD on copper films~\cite{Lee:2015}. 
These advances are an important step towards the realization of high-performance, large-scale graphene-based materials and devices.

The advent of graphene also spurred the study of other novel 2D atomic crystals, such as hexagonal boron nitride (h-BN), molybdenum disulfide, tungsten diselenide, and many others~\cite{Xu:2013,Gupta:2015}.
Like graphene, these 2D materials can be obtained by methods such as mechanical exfoliation, CVD growth, and epitaxial growth. 
Many 2D materials have distinct properties that are suited for practical applications in nanophotonic and quantum photonic devices~\cite{Shiue:2017}. 
For instance, monolayer h-BN is an optically transparent insulator, while monolayer $\mathrm{MoS_2}$ and $\mathrm{WSe_2}$ are direct band gap semiconductors with a strong photoluminescence~\cite{Xu:2013,Gupta:2015}. 
Large-area high-quality monolayer h-BN films with a size of $2\times 5~\mathrm{cm}$ have recently been grown by CVD on platinum films, where the size is limited only by the platinum foil size~\cite{ParkPark:2014}.
Importantly, these 2D materials, including graphene, can be reassembled layer by layer in a chosen sequence by mechanical stacking or CVD growth to form new 3D materials with diverse properties and atomically precise interfaces~\cite{Geim:2013,Niu:2015}. 
In the resulting so-called van der Waals (vdW) heterostructures, strong covalent bonds provide in-plane stability of 2D crystals, while relatively weak, vdW-like forces are sufficient to keep the stack together with an interlayer spacing of ${\lesssim\!1}$~nm. 
We highlight that this feature renders vdW heterostructures mechanically far more stable than ssDNA arrays. This is because in the latter the interstrand interactions are due to hydrogen bonds, which are much stronger than vdW-like forces and hence tend to entangle neighboring strands.
Recently, atomically thin photodetectors (or phototransistors) sensitive to light from the infrared (IR) to the ultraviolet (UV) have been fabricated using various vdW heterostructures~\cite{Koppens:2014}. 
With intrinsic mechanical stability, integrated designer optoelectronic functionality, and steady improvement in large-area, high-quality fabrication techniques, it is evident that vdW heterostructures are superior to ssDNA arrays as the DM detection material. Indeed, graphene layers and carbon nanotubes have been suggested as targets for directional detection of DM~\cite{Hochberg:2016ntt,Capparelli:2014lua,Cavoto:2016lqo,Cavoto:2017otc,Baracchini:2018wwj} and direct detection of cosmic neutrino background~\cite{Baracchini:2018wwj,Betti:2019ouf}.  

\section{Conceptual Design of the Proposed Detector}\label{sec:design}

\subsection{Detector configuration}\label{subsec:detector}

\begin{figure}[t]
\begin{center}
\includegraphics[width=3.25in]{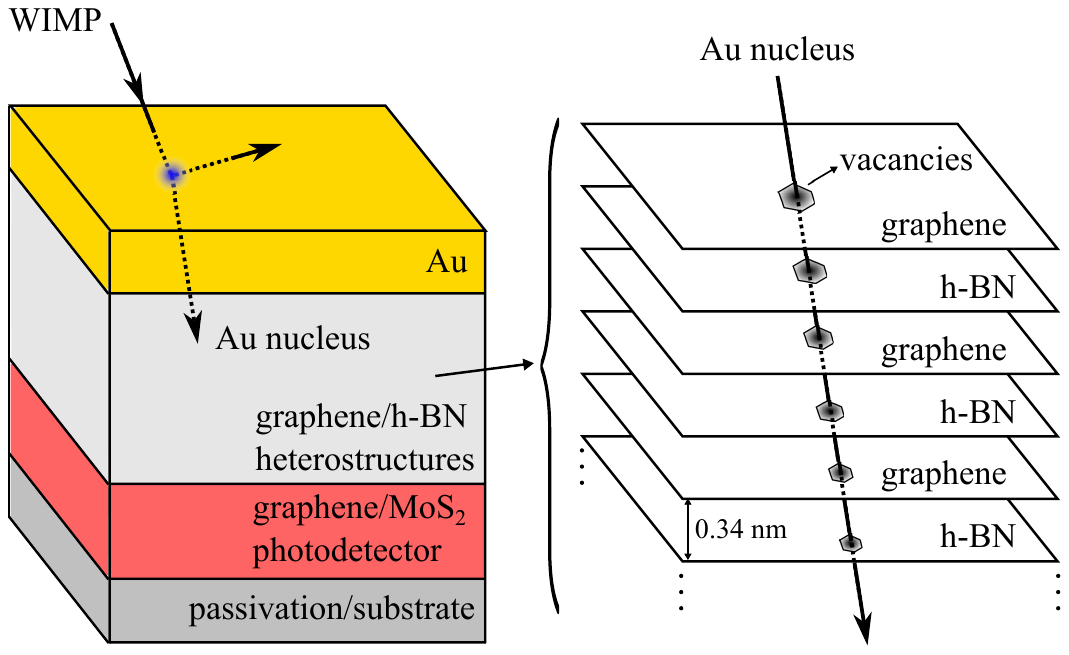}
\end{center}
\caption{(left) Schematic illustration of a single ``cell'' of the proposed detector (not to scale in the vertical direction). A WIMP scatters elastically off a Au nucleus in the gold film, and the recoiling nucleus traverses the graphene/h-BN heterostructure. (right) Schematic close-up of the graphene/h-BN heterostructure and the primary vacancies (shaded areas) created by the recoiling Au nucleus. For the purpose of clarity, secondary collisions and vacancies are not shown. The number, density, and size of the vacancies (represented by the size of the shaded area) provide information for intrinsic head-tail discrimination of the recoil track.}
\label{fig:figure1}
\end{figure}

We now present the conceptual design of the proposed graphene-based detector that leverages currently available materials and technology. It is noted that the proposed detector is in sharp contrast with a recently suggested graphene detector for sub-$\mathrm{GeV}/c^2$ WIMPs~\cite{Hochberg:2016ntt}, which utilizes electron recoils and is a 2D variant of conventional solid-state time-of-flight detectors with semiconductor targets replaced by graphene sheets as the target material.
The proposed detector consists of pixelated sheets of a thin film of metal, a graphene/h-BN heterostructure, a graphene/$\mathrm{MoS_2}$ heterostructure, and a substrate layer.
Figure~\ref{fig:figure1} depicts a schematic illustration of a single ``cell'' (or module) of the detector.
Thanks to modularity in the design, analogous to fabrication of integrated circuits, a large-size multilayered detector array can be built by stacking vertically and juxtaposing laterally a large number of single cells. Each cell consists of four components, which will now be described in detail.

\begin{figure}[t]
\begin{center}
\includegraphics[width=3.25in]{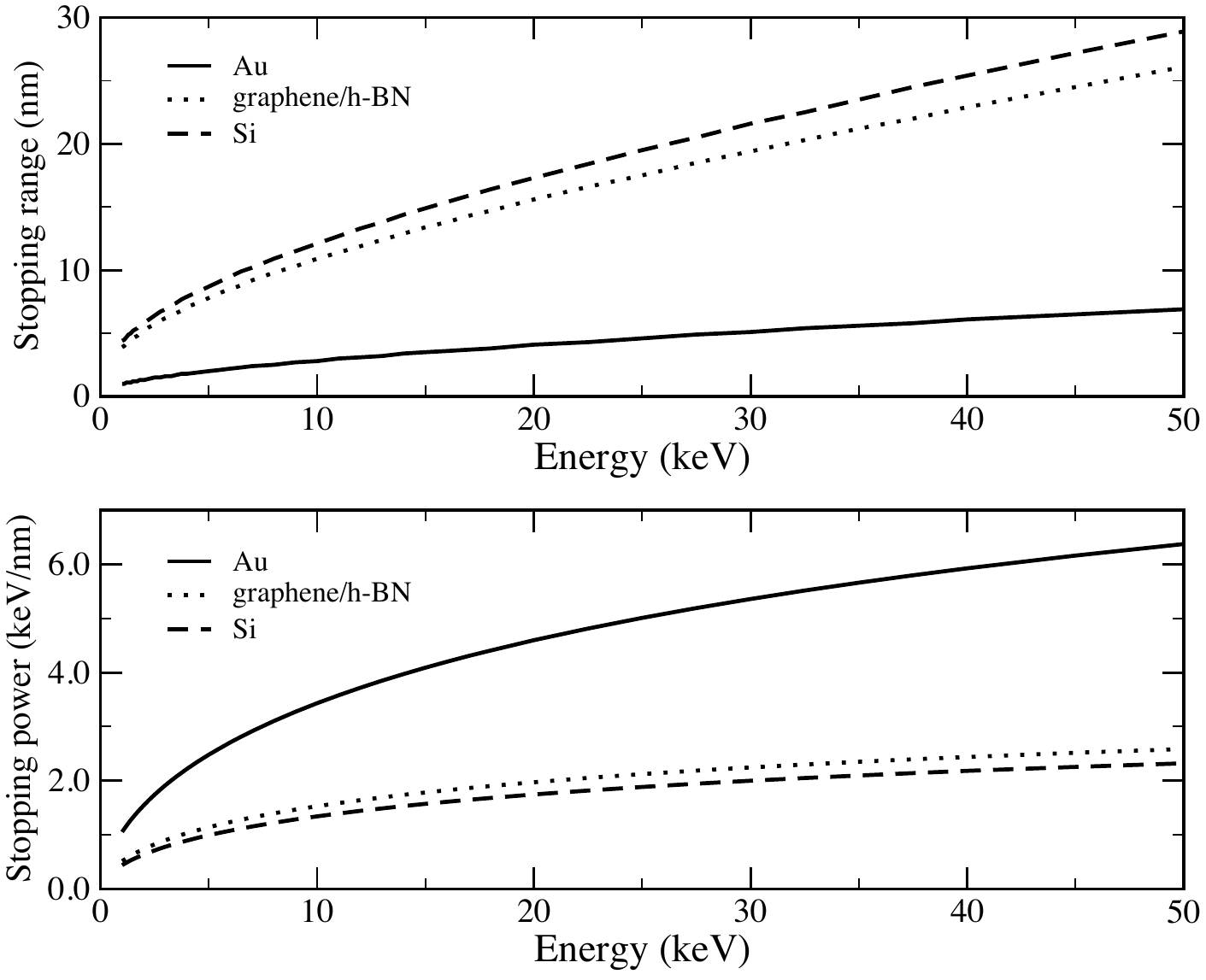}
\end{center}
\caption{Plot of SRIM simulated stopping range (top) of, and stopping power (bottom) for, a Au nucleus in Au, graphene/h-BN heterostructure, and Si as a function of Au nucleus energy. The simulations for the graphene/h-BN heterostructure were done by approximating the heterostructure by a compound of C, B, and N with a stoichiometric ratio of ${2\ratio 1\ratio 1}$.}
\label{fig:figure2}
\end{figure}

\begin{itemize}
\item[(i)]
The \emph{target} component is a thin film of metal. The choice of metal depends on the mass of the WIMPs to be detected, and its thickness is roughly 10--20 atoms thick so as to reduce multiple scattering of the recoiling nucleus in the film. Following Ref.~\cite{Drukier:2012hj}, with an incoming WIMP of mass ${\sim\!100}~\mathrm{GeV}/c^2$ in mind, here we use a 5-nm-thick gold film as the target.
\item[(ii)]
The \emph{detection} component is a graphene/h-BN heterostructure~\cite{Zhang:2015} with a thickness of about 7~nm. 
The distance between neighboring boron and nitrogen atoms in h-BN is 0.145~nm, which is only slightly larger than that of 0.142~nm between neighboring carbon atoms in graphene~\cite{Park:2014}, and 
a graphene sheet on a h-BN surface has an interlayer spacing of about 0.34~nm~\cite{Giovannetti:2007}. 
When a recoiling Au nucleus collides with and knocks out atoms in the graphene/h-BN heterostructure, vacancy defects~\cite{Hashimoto:2004,Banhart:2011,Robertson:2013} are created and a cascade of secondary collisions is produced in the heterostructure.
The total thickness of the graphene/h-BN heterostructure is chosen such that they are sufficiently thick but remain relatively optically thin in the visible-UV region. 
The frequency-dependent optical absorbance of graphene from the IR to the UV regions has been measured~\cite{Mak:2011}. 
It takes the universal value $\pi\alpha\approx 2.3\%$ (where $\alpha\approx 1/137$ is the fine structure constant)~\cite{Nair:2008} and is nearly frequency independent in the IR region of 0.5--1.5~eV, then increases smoothly and steadily to ${\sim\!4.1}\%$ at 3.0~eV in the visible region, and finally more rapidly to ${\sim\!5.8}\%$ up to 4.5~eV in the UV region~\cite{Mak:2011}. 
In contrast, h-BN is optically transparent in the same spectral regions because of a sufficiently large optical band gap of 6.05~eV~\cite{Wen:2015}. 
With the optical absorbance of graphene in the visible-UV region, we use a total of about 20 graphene and h-BN monolayers, corresponding to 7~nm in thickness.
The monolayers of the graphene/h-BN heterostructure can be aligned by doping impurities (e.g., using ion irradiation~\cite{Wang:2012}) in the corner region of the monolayers during the fabrication process. 
Specifically, two distinct dopants (e.g., arsenic and gallium) doped at two fixed adjacent corners of each monolayer in the heterostructure with a doping size of about $3\times 3~\mathrm{nm}$ allows for an alignment with nanometer precision.

\item[(iii)]
The \emph{registration} component is a graphene/$\mathrm{MoS_2}$ heterostructure photodetector. Monolayer $\mathrm{MoS_2}$ has a direct band gap of 1.8~eV~\cite{Zhang:2014}. A $\mathrm{MoS_2}$ monolayer is about 0.65~nm thick, and a graphene/$\mathrm{MoS_2}$ bilayer is about 1~nm thick. Including electrical contacts and the $\mathrm{SiO_2}$/Si passivation/substrate layer, the overall thickness of the registration component is about 200~nm~\cite{Koppens:2014,Zhang:2014}. 
With current technology, under laser illumination such a photodetector can reach a maximum internal quantum efficiency of about ${\sim\!15}$\% and a photoconductive gain up to $10^8$ at room temperature in the visible-UV region of 1.8--4.5~eV~\cite{Zhang:2014}. 
The internal quantum efficiency is defined as the number of photoexcited electron-hole pairs divided by the number of absorbed photons, and the photoconductive gain as the number of detected charge carriers divided by the number of photoexcited electron-hole pairs~\cite{Zhang:2014}.
The graphene/$\mathrm{MoS_2}$ photodetector is used to detect visible and UV bremsstrahlung photons~\cite{Astapenko:2013} produced by a recoiling Au nucleus and an ensuing cascade of secondary collisions in the graphene/h-BN heterostructure of the detection component. 
We stress that the registration component should \emph{not} be confused with the detection component \emph{nor} with the readout device, which will be specified in the following subsection. 
\item[(iv)]
The Si substrate of the graphene/$\mathrm{MoS_2}$ photodetector also serves as the \emph{absorption} component.
Its purpose is (a) to stop the recoiling nucleus and confine the collision products to its originating cell, and (b) to block out recoiling nuclei and collision products that originate outside the cell.
Combining the four components, we arrive at a total thickness of about 212~nm for a single cell in our conceptual design.
\end{itemize}
Figure~\ref{fig:figure2} shows the SRIM~\cite{SRIM} simulated stopping range of, and stopping power for, a Au nucleus in Au, graphene/h-BN heterostructure, and Si as a function of Au nucleus energy.
The Au thin film and the graphene/h-BN and graphene/$\mathrm{MoS_2}$ heterostructures are fabricated layer by layer on the substrate using CVD growth or mechanical stacking, and the substrate is also equipped with nanostructured electrical contacts.
Based on state-of-the-art fabrication technology~\cite{Lee:2015,ParkPark:2014,Wen:2015,Ma:2015}, the area of graphene and h-BN monolayers can reach ${\sim\!100}~\mathrm{cm}^2$, which also determines the area of a single cell. 
Currently the area of graphene/$\mathrm{MoS_2}$ photodetectors is about ${\sim\!50}~\mathrm{mm}^2$~\cite{DeFazio:2016}, and hence there are about 200 photodetectors in each cell, all fabricated on the same substrate layer so as to cover the area of a single cell.
For a 1-kg target of gold, a 5-nm-thick gold film has a surface area of $10^8~\mathrm{cm}^2$, corresponding to $10^6$ cells, each with an area of $100~\mathrm{cm}^2$. 
The detector has a volume of ${\sim\!2\times10^{-3}}~\mathrm{m}^3$, and contains 1.0~kg of Au, 90.7~g of graphene, 73.5~g of h-BN, 25.3~g of $\mathrm{MoS_2}$, and 4.8~kg of $\mathrm{SiO_2}$/Si substrate. 
Advances in manufacturing technology have enabled fabrication of flash memory in which 96 layers of memory cells are stacked vertically~\cite{Ahn:2018}.
Suppose that about 100 vertically stacked cell layers are fabricated on a standard 300~mm (12~inch) 775-$\mu$m-thick silicon wafer, then the entire detector has a volume of $\sim0.1~\mathrm{m}^3$ and contains an additional 220~kg of silicon. 
The latter numbers may be further reduced using the newly developed three-dimensional integrated circuit stacking technology~\cite{Shulaker:2017}.

\subsection{Detection principle}\label{sec:detection}

The detection principle of the proposed detector is based on defect formation produced by the recoiling nucleus in the graphene/h-BN heterostructure, and on visible and UV bremsstrahlung generated by the recoiling nucleus and a cascade of secondary collisions.
Vacancies in monolayer graphene and h-BN created by ion and electron irradiation have been under extensive experimental study~\cite{Hashimoto:2004,Banhart:2011,Robertson:2013,Meyer:2012,Kotakoski:2010}, and ion irradiation has recently used as an atomic doping technique for graphene~\cite{Wang:2012}. 
Bremsstrahlung produced by charged particles in nanostructures and solids is a burgeoning research topic in modern atomic physics~\cite{Shanker:2006,Tian:2009,Astapenko:2012,Astapenko:2013}.

An incoming WIMP from the Galactic halo collides elastically with one of the gold nuclei and knocks it out of the film. To be quantitative, following Refs.~\cite{Goodman:1984dc,Lewin:1995rx}, we consider a WIMP of mass $m$ scatters elastically off a target nucleus of mass $M$. The recoil energy of the nucleus is
\begin{equation}
E=\frac{v^2\mu^2}{M}(1-\cos\theta),
\end{equation}
where $\mu=mM/(m+M)$ is the reduced mass of the WIMP-nucleus system, $v$ is the speed of the WIMP relative to the nucleus, and $\theta$ is the scattering angle in the center of mass frame.
The recoil momentum of the nucleus is at most $2\mu v$ and the maximum recoil energy is 
\begin{equation}
E_\text{max}=\frac{2\mu^2v^2}{M}.\label{eq:Emax}
\end{equation}
We assume a typical recoil energy $E=E_\text{max}/2=\mu^2v^2/M$~\cite{Drukier:2012hj} and use the standard halo model~\cite{McMillan:2010} with $v_0=220~\mathrm{km}/\mathrm{s}$ for the mean WIMP velocity. 
For a WIMP of mass $m=200~\mathrm{GeV}/c^2$, the typical recoil energy of a Au nucleus is $E\sim 30$~keV. Specifically, we consider the situation that the WIMP-Au nucleus scattering takes place at the midpoint of 2.5~nm from the surface of the gold film. 
According to SRIM simulations (see Fig.~\ref{fig:figure2}), at this energy the stopping range of a Au nucleus in a gold target is 5.1~nm, and the stopping power of the target for the nucleus is $5.36~\mathrm{keV}/\mathrm{nm}$. Hence, a typical recoiling Au nucleus does escape from the gold film, and its energy after the escape is roughly halved, at 16.6~keV.

\begin{figure}[t]
\begin{center}
\includegraphics[width=3.25in]{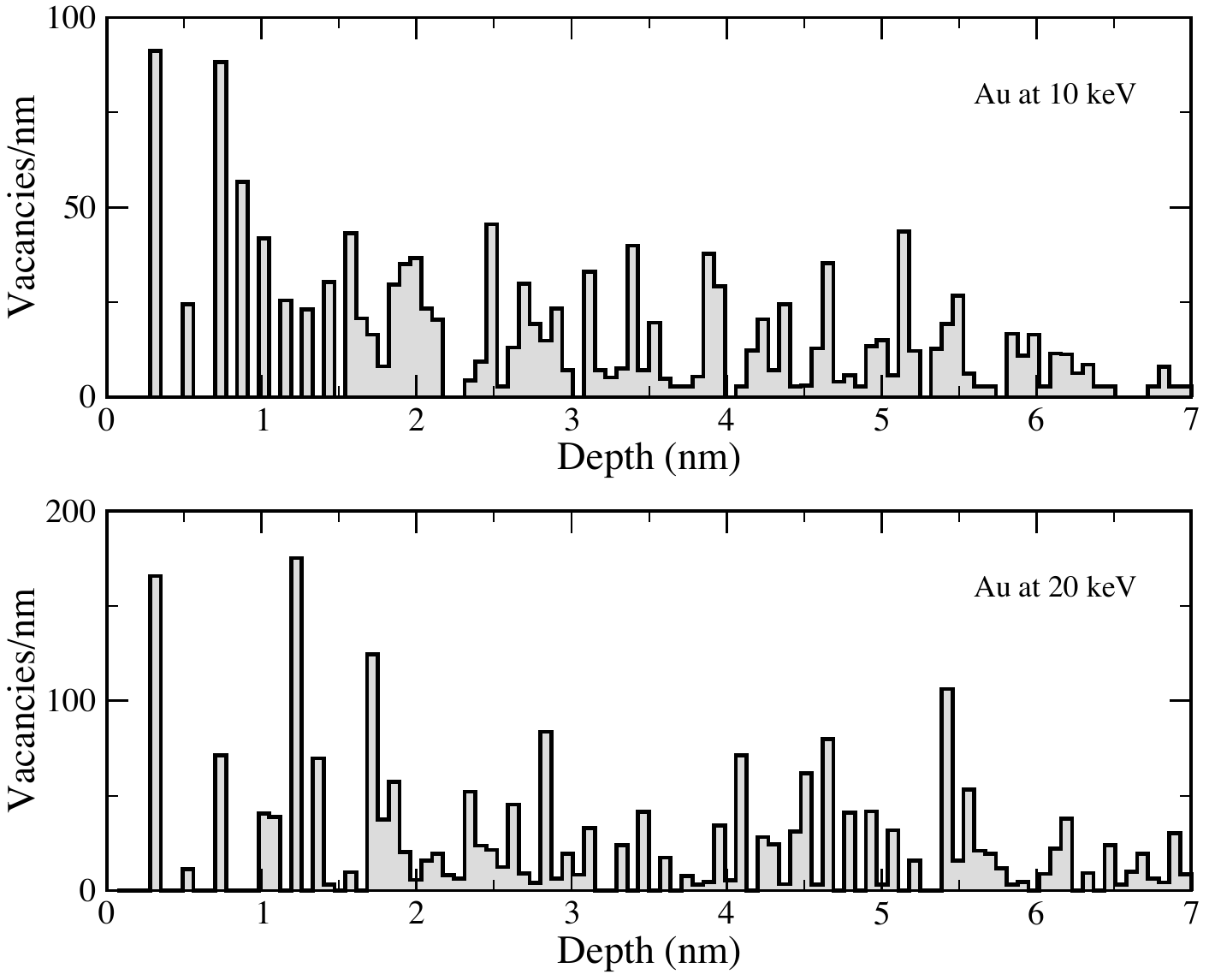}
\end{center}
\caption{Plot of SRIM simulated number of primary and secondary vacancies per nm of depth in the graphene/h-BN heterostructure as a function of recoiling Au nucleus depth in the heterostructure. The vacancies are produced by a recoiling Au nucleus entering at 10 (upper) and 20~keV (lower) in the direction perpendicular to the heterostructure. The simulations were done by approximating the graphene/h-BN heterostructure by a 7-nm-thick layer of C, B, and N with a stoichiometric ratio of ${2\ratio 1\ratio 1}$. }
\label{fig:figure3}
\end{figure}

The recoiling Au nucleus traverses and, whenever it knocks out atoms in the graphene and h-BN monolayers, creates vacancy defects and produces a cascade of secondary collisions in the graphene/h-BN heterostructure before either coming to a stop or being captured in the substrate (see Fig.~\ref{fig:figure1}). 
The displacement threshold energies (i.e., the minimum amount of kinetic energy that, when transferred to an atom, results in ejection of the atom away from its lattice site) for carbon atoms in monolayer graphene~\cite{Meyer:2012} and for boron and nitrogen atoms in monolayer h-BN~\cite{Kotakoski:2010} have been measured.
For ejection perpendicular to the layer, the displacement thresholds of C, B, and N are 23.6, 19.36, and 23.06~eV, respectively. Moreover, molecular dynamics simulations~\cite{Zobelli:2007} have shown that the displacement thresholds increase gradually for oblique ejection and become almost doubled for ejection parallel to the layer. 
SRIM simulations were carried out for an estimate of the number of vacancies produced by the recoiling Au nucleus in the graphene/h-BN heterostructure.
The binding energies of C, B, and N in the simulations were set at the above experimentally determined displacement threshold values. 
For a recoiling Au nucleus entering the graphene/h-BN heterostructure at 10 and 20~keV, there are on average about 100 and 150 (primary plus secondary) vacancies created in the heterostructure, respectively. The results of SRIM simulations are plotted in Fig.~\ref{fig:figure3}, assuming that the direction of the recoiling Au nucleus is perpendicular to the heterostructure.
We note that the alternating arrangement of the graphene and h-BN monolayers in the heterostructure (see Fig.~\ref{fig:figure1}, right panel) has the advantage of inhibiting possible interlayer vacancy migration~\cite{Liu:2014} between neighboring layers of the same kind. 
From the SRIM simulation results (see also Fig.~\ref{fig:figure2}), we find that after leaving the graphene/h-BN heterostructure the energy of the recoiling Au nucleus is significantly reduced to 3.5~keV, and the Au nucleus will subsequently be stopped by the substrate layer of the graphene/$\mathrm{MoS_2}$ photodetector.

Meanwhile, the visible and UV bremsstrahlung photons produced by the recoiling Au nucleus and a cascade of secondary collisions in the graphene/h-BN heterostructure are picked up by the graphene/$\mathrm{MoS_2}$ photodetectors in the cell, giving rise to induced photocurrents.
There are two mechanisms of bremsstrahlung for an incident charged particle on an atom, i.e., ordinary and polarization (or atomic) bremsstrahlung~\cite{Astapenko:2013,Amusia:2006}. 
The former is caused by the deflection of the incident charged particle in the Coulomb field of the target particle; the latter arises from the dynamic polarization of the target electrons by the Coulomb field of the incident charged particle.
While bremsstrahlung is commonly associated with radiation in the X-ray region, visible and UV bremsstrahlung has been applied to study inertial confinement fusion~\cite{Chen:2015} and laser pulse heated dielectrics~\cite{Petrov:2017}. 
We note that UV and X-ray bremsstrahlung photons from nuclear recoils in conventional liquid xenon detectors have recently been suggested as possible signatures for sub-$\mathrm{GeV}/c^2$ WIMP direct detection~\cite{Kouvaris:2016afs}.
It has been shown~\cite{Amusia:2006,Korola:2006} that for a heavy charged projectile of mass $M_\mathrm{proj}\gg m_\mathrm{e}$, where $m_\mathrm{e}$ is the electron mass, polarization bremsstrahlung (PB) is more important than ordinary bremsstrahlung (OB) down to low photon energies 
\begin{equation}
\hbar\omega\gtrsim \frac{I}{\sqrt{(M_\mathrm{proj}/m_\mathrm{e})}},
\end{equation} 
where $\omega$ is the photon frequency and $I$ is the ionization energy of the target atom. 
The ionization energies of C, B, and N are about ${\sim\!10}$~eV, and hence for the recoiling Au nucleus we have $\hbar\omega\gtrsim 0.5$~eV.
Therefore, the dominant contribution to the Au nucleus bremsstrahlung in the visible-UV region comes from PB, which has an intensity independent of the projectile mass~\cite{Amusia:2006}.
The induced photocurrents are fed into a circuit connected to an external electronic device (not shown in Fig.~\ref{fig:figure1}) that analyzes the photocurrent profiles, registers the nuclear recoil event, and locating the corresponding cell and photodetector. 
This information allows for initial coarse localization of the recoil track within a ${\sim\!50}~\mathrm{mm}^2\times 7~\mathrm{nm}$ volume in the graphene/h-BN heterostructure of the registered cell. 
While photocurrent spectra (i.e., photocurrent vs photon energy curves) of few-layer $\mathrm{MoS_2}$ photodetectors under laser illumination have recently been measured over the photon energy range from 1.25 to 2.5~eV~\cite{Li:2017}, the sensitivity of $\mathrm{MoS_2}$ photodetectors to single photons is yet to be investigated.
Moreover, bremsstrahlung spectra from heavy ions and cascades of secondary collisions on graphene/h-BN heterostructures in the visible-UV region are not yet available. The same is also true for the photocurrent spectra of graphene/$\mathrm{MoS_2}$ photodetectors over the 2.5--4.5~eV photon energy range. Hence we will leave the analysis of those to future work.

\begin{figure}[t]
\begin{center}
\includegraphics[width=3.25in]{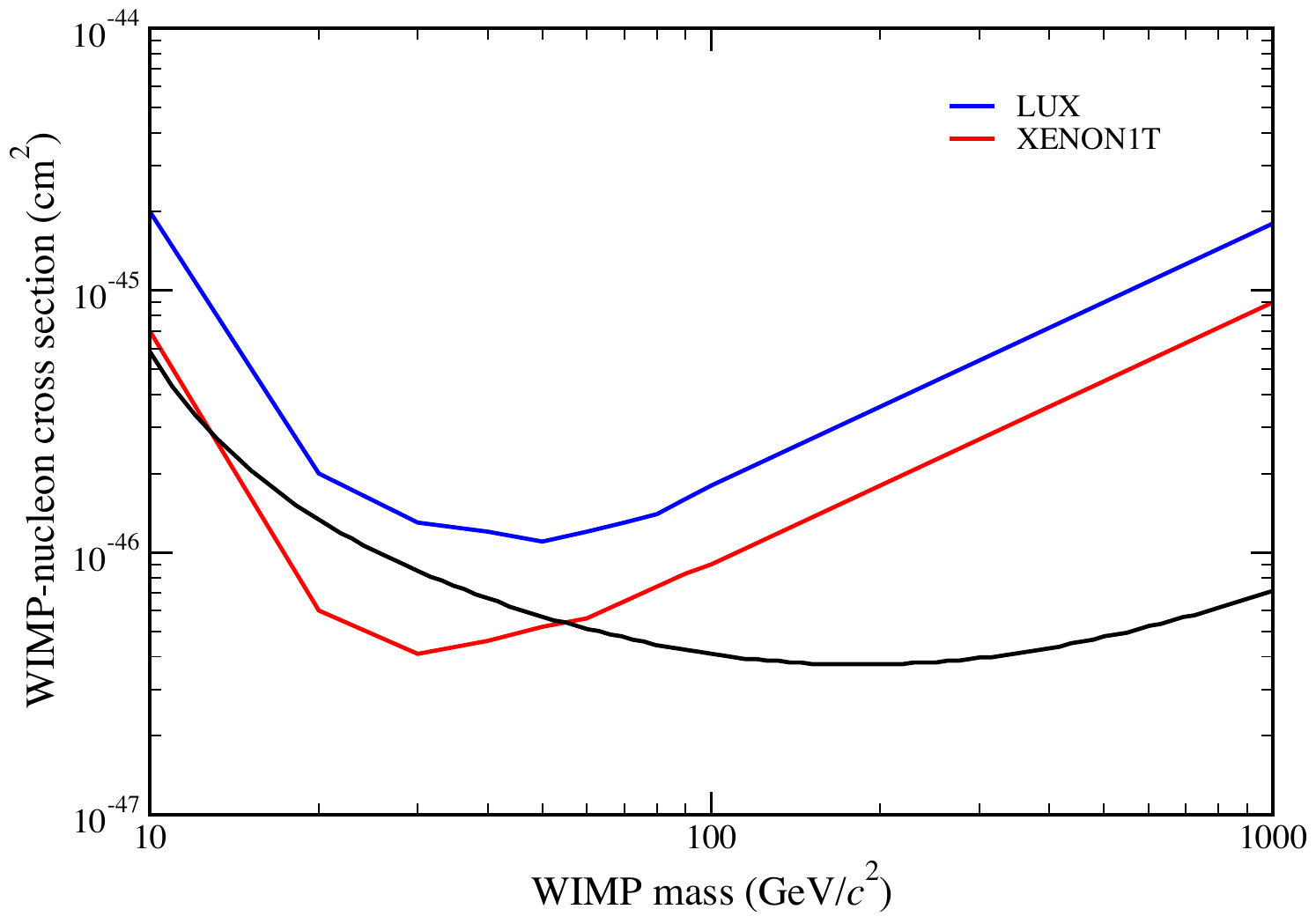}
\end{center}
\caption{Expected background-free 90\% confidence level sensitivity for a Au target with an expected 3.2 events and a 100-kg-year exposure (black line), assuming all the WIMP-induced events are identified, and exclusion limits from LUX~\cite{Akerib:2016vxi} (blue line) and XENON1T~\cite{Aprile:2018dbl} (red line) experiments.}
\label{fig:figure4}
\end{figure}

Recall that the 7-nm-thick graphene/h-BN heterostructure in one single cell consists of a total of about 20 graphene and h-BN monolayers. 
We define the energy threshold of the detector as the smallest energy for a recoiling nucleus to traverse one half the depth of the graphene/h-BN heterostructure. Then the energy threshold is about ${\sim\!1}$~keV, which is verified by SRIM simulations. 
This energy threshold is comparable to but lower than those (${\sim\!2}$--4~keV) of low-pressure gaseous TPC-based detectors~\cite{Ahlen:2009ev}.
The differential recoil rate per unit detector mass can be written as (typically given in units of events per keV per kg per day)~\cite{Bertone:2004pz,Lewin:1995rx}
\begin{equation}
\frac{dR}{dE}=\frac{\sigma_0\rho_0}{2m\mu^2}F^2(q)\int_{v>v_\mathrm{min}}\frac{f(\mathbf{v})}{v}d^3v,\label{eq:dRoverdE}
\end{equation}
where $\rho_0$ is the local dark matter mass density, $q=\sqrt{2ME}$ is the momentum transfer, $\sigma_0$ is the WIMP-nucleus cross section at zero momentum transfer, $F(q)$ is the nuclear form factor, $f(\mathbf{v})$ is the WIMP velocity distribution, and $v_\mathrm{min}=\sqrt{ME/2\mu^2}$ is the minimum WIMP velocity that can give rise to a recoil energy $E$.
For simplicity, we consider only the spin-independent cross section for elastic scattering of a WIMP with a nucleus. The spin-independent WIMP-nucleus cross section at zero momentum transfer is typically of the form
\begin{equation}
\sigma_0=\frac{\mu^2}{\mu_p^2}A^2\sigma_p,
\end{equation}
where $A$ is the atomic mass of the nucleus, $\mu_p$ is the WIMP-proton reduced mass, and $\sigma_p$ is the spin-independent cross section of a WIMP with the proton at zero momentum transfer. 
The integrated rate is then obtained by integrating Eq.~\eqref{eq:dRoverdE} over a recoil energy range $E_\mathrm{min}$ to $E_\mathrm{max}$, where $E_\mathrm{max}$ is the maximum recoil energy given by Eq.~\eqref{eq:Emax} and $E_\mathrm{min}$ is the minimum recoil energy corresponding to the energy threshold of the detector. 
From the SRIM result for the stopping power shown in Fig.~\ref{fig:figure2}, we find that for a WIMP-Au nucleus scattering taking place at the midpoint of 2.5 nm from the surface of the gold film an energy threshold of 1~keV corresponds to $E_\mathrm{min}\sim 8~\mathrm{keV}$. 
Moreover, with a suitable target material (e.g., beryllium) with appropriate target thickness, the low detector energy threshold of 1~keV also allows for directional detection of WIMPs down to ${\sim\!10}~\mathrm{GeV}/c^2$ mass.
Specifically, for a $10~\mathrm{GeV}/c^2$ WIMP and a 5-nm-thick beryllium film, SRIM simulations shows that the energy of a typical recoiling Be nucleus after escaping from the film is about 1.34~keV, and on average there are about 13 vacancies created in the graphene/h-BN heterostructure. 
We assume the standard halo model with local dark matter density $\rho_0=0.3~\mathrm{GeV}/\mathrm{cm}^3$, mean WIMP velocity $v_0=220~\mathrm{km}/\mathrm{s}$, and WIMP escape velocity $v_\text{esc}=550~\mathrm{km}/\mathrm{s}$~\cite{McMillan:2010}. 
In Fig.~\ref{fig:figure4}, we show the one-sided Poisson 90\% confidence level sensitivity for a Au target with an expected reach of 3.2 events after a 100-kg-year exposure, assuming a background-free experiment and all the WIMP-induced events are identified, together with the exclusion limits from LUX~\cite{Akerib:2016vxi} and XENON1T~\cite{Aprile:2018dbl} experiments. It is evident that the proposed detector with a Au target is capable of probing the $\sim\!50$--1000~$\mathrm{GeV}/c^2$ WIMP mass range in the WIMP-nucleon cross section vs WIMP mass parameter space that is currently not ruled out by direct detection experiments.

After an exposure during which an expected number of nuclear recoil events are registered, the registered cells can be removed and their component graphene/h-BN heterostructures be delaminated by mechanical exfoliation, while the gold films, graphene/$\mathrm{MoS_2}$ photodetectors, and substrate layers are recycled for reuse.
The position of the vacancies in the now graphene and h-BN monolayers can be determined with atomic precision by a separate readout device, such as the transmission electron microscope (TEM) or high-resolution TEM (HRTEM)~\cite{Hashimoto:2004,Banhart:2011,Robertson:2013,Wang:2012}. 
The information about the vacancies, together with the structure information about the graphene/h-BN heterostructure, allows the path of the recoiling Au nucleus to be tracked with nanometer precision.
The area of a single graphene or h-BN monolayer needs to be scanned by the TEM is about $50~\mathrm{mm}^2$, i.e., the area of the registered graphene/$\mathrm{MoS_2}$ photodetector in the initial localization of the recoil track. 
Assume using a TEM with a typical cell size of 1~\r{A} and typical dwell time of 1~ns for each cell. 
Then it takes about $2.5\times10^6~\mathrm{s}\approx 29~\mathrm{d}$ to sequentially scan the area and locate the vacancies in one graphene or h-BN monolayer, and there are about 20 graphene and h-BN monolayers to be scanned in one registered cell. 
Presumably, it takes about $29\times 20$~days $\approx 1.6$~years to scan a registered cell. However, it is important to note that we only want to locate vacancies, and that the positions of vacancies in adjacent monolayers are strongly correlated because the vacancies are created by a single recoiling nucleus.
Thus only the first monolayer (i.e., the one closest to the Au film) needs to be scanned sequentially until the vacancies are located. This is similar to the problem of searching for an element in an unsorted list, and hence it takes $29/2=14.5$ days on average and 29 days in the worst case to locate the vacancies.  
Once those vacancies are located, for the remaining 19 monolayers we only need to scan a small area that surrounds the position corresponding to the located vacancies in the previous monolayer. 
Therefore, the total scanning time for a registered cell is approximately 14.5~days on average and 29~days in the worst case, instead of 1.6~years.
Moreover, analogous to the idea of distributed computing, the scanning time of all registered cells can be further reduced using ``distributed scanning'' in the sense that the graphene and h-BN monolayers in the registered cells that need to be scanned can be distributed to, say, tens or a hundred of TEMs located possibly at different sites such that they can be scanned virtually simultaneously. 
Since only ${\sim\!10}$ recoil events will be sufficient to distinguish a WIMP signal from isotropic backgrounds, an efficient readout and track reconstruction is experimentally viable.

\subsection{Directionality and background rejection}\label{sec:directionality}

The proposed detector design has intrinsic head-tail discrimination and background rejection capabilities. We first note that in the proposed detector design if a recoiling Au nucleus is registered by the graphene/h-BN heterostructure, it always moves away from the Au film and traverses the graphene/h-BN heterostructure. 
Moreover, the number, density, and size (i.e., single vacancies with one atom missing, and multiple vacancies with two or more atoms missing) of vacancies created by a recoiling nucleus in the graphene/h-BN heterostructure directly depend on the energy of the recoiling nucleus~\cite{Banhart:2011,Wang:2012}. 
Because collisions of a recoiling nucleus with target atoms and atomic crystals dissipate energy, the recoiling nucleus always moves from layers with many, dense, and multiple vacancies to layers with few, sparse, and single vacancies.
Hence the direction of a recoil track (head-tail directionality) can be determined using the information obtained from TEM or HRTEM imaging about the number, density, and size of the vacancies in each graphene and h-BN monolayers (see also Fig.~\ref{fig:figure1}). 
Specifically, following Ref.~\cite{Rajendran:2017ynw}, we define the asymmetry of a track as the ratio of the number of vacancies in the end third of the recoiling Au nucleus depth in the heterostructure to the beginning third. 
The SRIM simulation results shown in Fig.~\ref{fig:figure3} indicates that there is a roughly $1\ratio 2$ asymmetry for a recoiling Au nucleus entering the graphene/h-BN heterostructure at 10 and 20~keV, supporting the head-tail discrimination capability of the detector.

While information about the directionality of the WIMP signal provides an inherent discriminant against backgrounds, the expected low event rate requires that the detector is built from radiopure materials, operated deep underground to shield against cosmic rays, and placed inside a water tank to shield out alphas, betas, gammas, and neutrons from the rock walls.
However, a background-free environment is virtually never attainable. The modular design with a dense high-$Z$ target material allows for effective volume fiducialization~\cite{Undagoitia:2015gya} by excluding events from cells in the outer regions of a large-size detector with a cubic geometry. 
This fiducialization is especially useful in suppressing neutron-induced background events that cannot be distinguished from a WIMP-induced signal.
The primary backgrounds are gammas from the environment and from decays within the detector. 
Here we focus on internal backgrounds that give rise to electron recoils in the detector.
According to the decay properties of the relevant beta emitters listed in Table~\ref{tab:radioisotopes}, we expect that the main irreducible backgrounds come from ${}^{14}\mathrm{C}$ decay in graphene and ${}^{100}\mathrm{Mo}$ decay in $\mathrm{MoS_2}$. 
On the one hand, accelerator mass spectroscopy during the fabrication process of graphene can reduce the ${}^{14}\mathrm{C}/\mathrm{C}$ ratio to $10^{-21}$~\cite{Litherland:2005}. For a detector with a 1-kg target of gold, this corresponds to ${\sim\!4.5\times 10^3}$ atoms of ${}^{14}\mathrm{C}$ in the graphene (see Sec.~\ref{subsec:detector}). 
With the ${}^{14}\mathrm{C}$ half-life of 5730 years, there are roughly 0.8 events per year assuming no veto.
On the other hand, anion-exchange chromatography can be used to purify the abundance of the most common stable isotope ${}^{98}\mathrm{Mo}$ to $99.9\%$~\cite{Malinovsky:2014}, corresponding to a ${}^{100}\mathrm{Mo}/\mathrm{Mo}$ ratio of $10^{-4}$ or ${\sim\!6.2\times 10^{18}}$ atoms of ${}^{100}\mathrm{Mo}$ in $\mathrm{MoS_2}$. With a half-life of $9.5\times10^{18}$ years for ${}^{100}\mathrm{Mo}$, this gives rise to 1.1 events per year, again assuming no veto.

\begin{table}
\begin{center}
\caption{Relevant beta emitters.\label{tab:radioisotopes}}\vspace{1ex}
\begin{tabular}{lllll}
\hline\hline
Isotope~~~& Abundance~~~& Half-life & Mode~~~&$Q$-value (MeV)\\\hline
${}^{3}\mathrm{H}$&trace&12.32~y&$\beta^-$&0.01861\\
${}^{14}\mathrm{C}$&trace&5730~y&$\beta^-$&0.156\\
${}^{31}\mathrm{Si}$&trace&2.62~h&$\beta^-$&1.495\\
${}^{32}\mathrm{Si}$&trace&153~y&$\beta^-$&13.020\\
${}^{32}\mathrm{P}$&trace&14.29~d&$\beta^-$&1.709\\
${}^{35}\mathrm{S}$&trace&87.32~d&$\beta^-$&0.167\\
${}^{100}\mathrm{Mo}$&9.63\%&${8.56\times10^{18}}$~y~~~&$\beta^-\beta^-~~$&3.034\\\hline\hline
\end{tabular}
\end{center}
\end{table}

\begin{figure}[t]
\begin{center}
\includegraphics[width=3.25in]{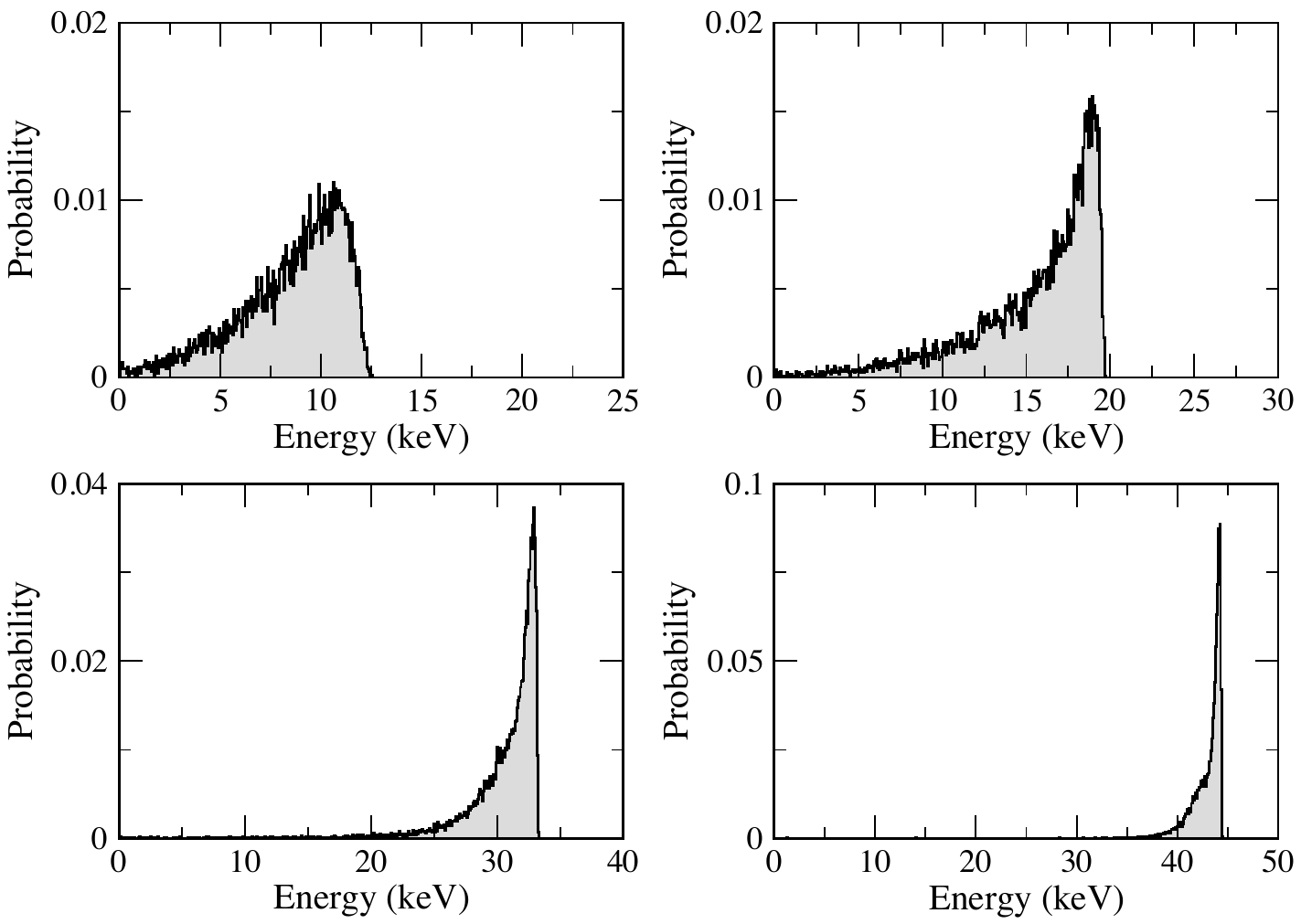}
\end{center}
\caption{Plot of CASINO simulated energy distribution of a transmitted electron traversing a single cell of the detector (from Au to substrate) at 25 (upper left), 30 (upper right), 40 (lower left), and 50~keV (lower right). The simulations were done by approximating the graphene/h-BN heterostructure by a 7-nm-thick layer of C, B, and N with a stoichiometric ratio of ${2\ratio 1\ratio 1}$, and the graphene/$\mathrm{MoS_2}$ photodetector by a 1-nm-thick layer of C, Mo, and S with a stoichiometric ratio of ${3\ratio 1\ratio 2}$.}
\label{fig:figure5}
\end{figure}

The recoiling electrons undergo scattering and bremsstrahlung in the detector.
CASINO Monte Carlo simulations~\cite{CASINO} were carried out for an assessment of  electron stopping in a single cell of the detector. Simulation results for the energy distribution of a transmitted electron are plotted in Fig.~\ref{fig:figure5}. 
For the internal backgrounds under consideration, the results imply that (i) recoiling electrons at ${\gtrsim\!10}$~keV are most likely to traverse multiple cells, thus producing time-coincident photocurrents in neighboring cells that can subsequently be rejected, and (ii) recoiling electrons at ${\lesssim\!10}$~keV will be stopped and trapped inside a single cell, and generate photocurrents in the cell.
Therefore, electron recoils from ${}^{14}\mathrm{C}$ and ${}^{100}\mathrm{Mo}$ decays at ${\gtrsim\!10}$~keV can be easily rejected by coincidence counting because of their multicell interaction.
To estimate the event rate for electron recoils at ${\lesssim\!10}$~keV, we use the measurement results for the ${}^{14}\mathrm{C}$ beta energy spectrum in the energy range of 10--160~keV~\cite{Kuzminov:2000up} and the ${}^{100}\mathrm{Mo}$ double-beta energy sum spectrum in the 0.4--3.2~MeV range~\cite{Arnold:2005rz}.
Extrapolating the results down to ${\sim\!10}$~keV, we find that the probability of the ${}^{14}\mathrm{C}$ decay electron with energy less than 10~keV is about 0.1, and the probability of the ${}^{100}\mathrm{Mo}$ decay electron pair with total energy less than 20~keV is approximately upper bounded by $10^{-4}$. 
Thus we obtain 0.08 electron recoil events per year at ${\lesssim\!10}$~keV, which is largely negligible after taking into account of the small electron OB cross section compared with its nuclear PB counterpart (see below). 
The result also emphasizes the importance of achieving a ${}^{14}\mathrm{C}/\mathrm{C}$ radiopurity level of $10^{-21}$ or better.
Electron recoils at ${\lesssim\!10}$~keV will leave tracks in the graphene/h-BN heterostructure and there is still a chance, albeit small, for those electron recoils to be registered. Their directions can be determined similarly to that of the nuclear recoils. 
However, as noted in the Introduction, since electrons recoils caused by random backgrounds are expected to be isotropic and exhibit no preferred direction, according to the analysis of Ref.~\cite{Morgan:2004ys} only ${\sim\!10}$ recoil events will be sufficient to distinguish a WIMP-induced event from the background events.

Finally, we provide an estimate of the bremsstrahlung cross sections of an electron and a Au nucleus on monolayer graphene in the visible-UV region.
Studies of bremsstrahlung on 2D materials are rather scarce, although some progress has been made in recent years in the X-ray region. 
In Refs.~\cite{Astapenko:2012,Astapenko:2013} bremsstrahlung of 30 and 58~keV electrons on graphene has been studied in the photon energy range of 0.2--10~keV.
The results show that the differential cross sections for both OB and PB of an electron increase almost monotonically (except for some sharp peaks) with decreasing photon energy.
Note that the analysis for the dominance of nucleus PB at low photon energies discussed in Sec.~\ref{sec:detection} also applies to electron PB down to low photon energies $\hbar\omega\gtrsim I$~\cite{Amusia:2006,Korola:2006}. 
Since $I\approx 10$~eV for C, B, and N, the dominant contribution to electron bremsstrahlung in the visible-UV region comes from OB.
Using the fact that the electron OB cross section for a nonrelativistic electron scales approximately as the inverse square of the electron velocity, and extrapolating the electron OB results of Refs.~\cite{Astapenko:2012,Astapenko:2013} to the visible-UV region, we find that the OB differential cross section for an ${\lesssim\!20}$~keV electron on graphene is about ${\sim\!10^{-24}}~\mathrm{cm}^2/\mathrm{eV}$.
Moreover, because the differential cross section for PB is independent of the projectile mass and scales approximately as the inverse square of the projectile velocity, the Au nucleus PB differential cross section can be extrapolated using the electron PB results of Refs.~\cite{Astapenko:2012,Astapenko:2013}. 
The PB differential cross section per atom in the visible-UV region for an ${\lesssim\!20}$~keV Au nucleus on graphene is estimated to be about ${\sim\!10^{-22}}~\mathrm{cm}^2/\mathrm{eV}$, which is two orders of magnitude larger than its electron OB counterpart.
Moreover, the small electron OB cross section also justifies the neglect of electron recoils at ${\lesssim\!10}$~keV.

Several comments are in order.
First, we stress that the PB cross section estimated above is at best a lower bound of the Au nucleus PB cross section on the graphene/h-BN heterostructure. 
This is because the result of Refs.~\cite{Astapenko:2012,Astapenko:2013} is obtained in the hard photon limit $\hbar\omega\gg I\approx 10$~eV for monolayer graphene; hence soft photon emission as well as collective and interfacial effects in the graphene/h-BN heterostructure that may enhance the polarizability are not taken into account. 
Second, we note that the cascade of secondary collisions in the graphene/h-BN heterostructure will also produce visible and UV photons. 
Given the multiplicity of secondary collisions, their contributions to visible and UV photon production could be comparable to or even greater than that of the recoiling Au nucleus.
Third, it is interesting to consider integration of technologies such as graphene-based single-photon detection~\cite{Williams:2016} and 2D material-based single-photon emission~\cite{Aharonovich:2016} into the detector design so as to improve visible and UV photon production and detection.
In particular, polarized and ultrabright single-photon emission across the visible and the near-IR regions from vacancy defects in h-BN monolayers and multilayers has recently been observed under electron irradiation at room temperature~\cite{Tran:2016a,Tran:2016b}. 
The defect-induced photon emission from h-BN may significantly enhance the visible and UV photon intensities.
We hope to report on some of these aspects in the near future.

\section{Conclusions}\label{sec:conclusion}

We have proposed and studied a novel dark matter detector with nanometer precision and directional sensitivity using graphene-based van der Waals heterostructures, such as graphene/h-BN and graphene/$\mathrm{MoS_2}$ heterostructures. 
The proposed detector has modular scalability, keV-scale detection threshold, nanometer position resolution, sensitivity down to dark matter of ${\sim\!10}~\mathrm{GeV}/c^2$ mass, and intrinsic head-tail discrimination and background rejection capabilities.
The feasibility of the proposed design is strongly supported by a great amount of experimental evidence on the properties and fabrication of the heterostructures, and by simulation results.
It is fascinating that graphene-based 2D materials, heterostructure devices, and nanotechnology may help solve the old problem of dark matter. 
We leave for future work the studies of bremsstrahlung in the visible-UV region from heavy ions and cascades of secondary collisions on graphene/h-BN heterostructures, and the improvement and optimization of the proposed detector.

\acknowledgments

I would like to thank C.N.\ Leung and D.C.\ Ling for useful discussions and a careful reading of an early version of the manuscript.

\end{document}